\documentclass[sn-nature,iicol]{sn-jnl}
\usepackage{amsmath}
\usepackage[dvipsnames]{xcolor}         

\begin{document} 


\title{Detection of Single-Mode Thermal Microwave Photons Using an Underdamped Josephson Junction}

\author*[1,2]{\fnm{A.~L.} \sur{Pankratov}}
\email{alp@ipmras.ru}
\author*[1]{\fnm{A.~V.} \sur{Gordeeva}}\email{a.gordeeva@nntu.ru}
\author[1,2]{\fnm{A.~V.} \sur{Chiginev}}
\author[1,2]{\fnm{L.~S.} \sur{Revin}}
\author[1,2]{\fnm{A.~V.} \sur{Blagodatkin}}
\author[3,4]{\fnm{N.} \sur{Crescini}}
\author[1,5]{\fbox{\fnm{L.~S.} \sur{Kuzmin}}}



\affil[1]{\orgname{Nizhny Novgorod State Technical University n.a. R.~E. Alekseev},  \city{Nizhny Novgorod} \postcode{603950}, \country{Russia}}
\affil[2]{\orgname{Institute for Physics of Microstructures of RAS}, \city{Nizhny Novgorod} \postcode{603950}, \country{Russia}}
\affil[3]{\orgname{Fondazione Bruno Kessler (FBK)}, \orgaddress{\postcode{I-38123}, \city{Trento}, \country{Italy}}}
\affil[4]{\orgname{Univ. Grenoble Alpes, CNRS, Grenoble INP, Institut N\'eel}, \orgaddress{\postcode{38000} \city{Grenoble}, \country{France}}}
\affil[5]{\orgname{Chalmers University of Technology}, \orgaddress{\postcode{41296} \city{Gothenburg}, \country{Sweden}}}

\abstract{
    When measuring electromagnetic radiation of frequency $f$, the most sensitive detector is the one that counts the single quanta of energy $h f$. Single photon detectors (SPDs) were demonstrated from $\gamma$-rays to infrared wavelengths \cite{Natarajan2012Apr, Walsh2021Apr,PhysRevApplied.19.034007}, and extending this range down to the microwaves is the focus of intense research \cite{Astafiev2002Jun,Johnson2010Sep,Inomata2016Jul,Kono2018Jun,Lescanne2020May,Albertinale2021,Besse2018Apr}.
    The energy of 10\,GHz microwave photon, about $40\,\mu\text{eV}$ or $7\, \text{yJ},$ is enough to force a superconducting Josephson junction into its resistive state, making it suitable to be used as a sensor \cite{Oelsner2013Sep,Oelsner2017Jan,Chen2011Nov,Poudel2012Nov,Kuzmin2018Jun,Revin2020Jun,Pankratov2022May}. In this work, we use an underdamped Josephson junction to detect single thermal photons stochastically emitted by a microwave copper cavity  at millikelvin temperatures.
    After characterizing source and detector, we vary the temperature of the resonant cavity and measure the increased photon rate. The device shows an efficiency up to 45\% and a dark count rate of 0.1\,Hz in a bandwidth of several GHz. To confirm the thermal nature of the emitted photons we verify their super-Poissonian statistics \cite{Fox2006,Kovalenko:23}, which is also a signature of quantum chaos \cite{Yusipov2020Feb}.
    We discuss detector application in the scope of Dark Matter Axion searches \cite{Lamoreaux2013Aug}, and note its importance for quantum information \cite{PhysRevLett.102.173602,PhysRevA.90.062307,doi:10.1126/science.aat4625,Grimsmo2021Mar}, metrology \cite{newsi,Brida_2012} and fundamental physics \cite{doi:10.1146/annurev.nucl.012809.104433,Kuzmin2018Jun}.
}

\date{\today}


\maketitle 

\section*{Introduction}
Photons, the quanta of light, are massless particles whose energy $E$ is related to their frequency $f$ by the Planck's constant $h$ through the relation $E=h f$ \cite{RevModPhys.78.1267}.
The most energetic photons ever recorded have $E>100\,\text{TeV}$ \cite{PhysRevLett.123.051101} and belong to the class of gamma rays, which extends down to $E\sim\text{MeV}$. Lower energy photons are known as X-rays and have $E\sim\text{keV}$.
Single photons in the gamma and X-ray range are detected with techniques belonging to the world of high-energy physics, for example crystal scintillators and photomultiplier tubes \cite{Bettini2008}.
In the optical domain, a photon typically has an energy $E\sim\text{eV}$, and single photons can be detected by established means, for instance photomultipliers or avalanche photodiodes \cite{Hadfield2009Dec}.
The use of superconductors pushed the detectable energy down to the infrared range, with $E\sim\text{meV}$, by using nanowires \cite{EsmaeilZadeh2021May, Khasminskaya2016Nov}, transition-edge sensors \cite{Irwin2005Jul}, kinetic inductance detectors \cite{PhysRevApplied.19.034007} and Josephson junctions \cite{Walsh2021Apr}.
The microwave range, with $E\sim40\,\mu\text{eV}$ ($\sim\,7\, \text{yJ}$), is the frontier of single photon detectors \cite{Gleyzes2007Mar}, approached by quantum dots \cite{Astafiev2002Jun} and bolometric schemes \cite{Govenius2016Jul, Lee2020Oct, Kokkoniemi2020Oct}. 
The detection of single microwave photons attracted the attention of quantum technologies, and several circuit quantum electrodynamics schemes \cite{Blais2021May} were proposed \cite{Royer2018May, Grimsmo2021Mar} and implemented using qubits \cite{Johnson2010Sep,Inomata2016Jul,Kono2018Jun,Lescanne2020May,Albertinale2021,Dixit2021Apr,PhysRevApplied.21.014043,Besse2018Apr}.

\begin{figure}[!]
    \centering
    \includegraphics[width=.5\textwidth]{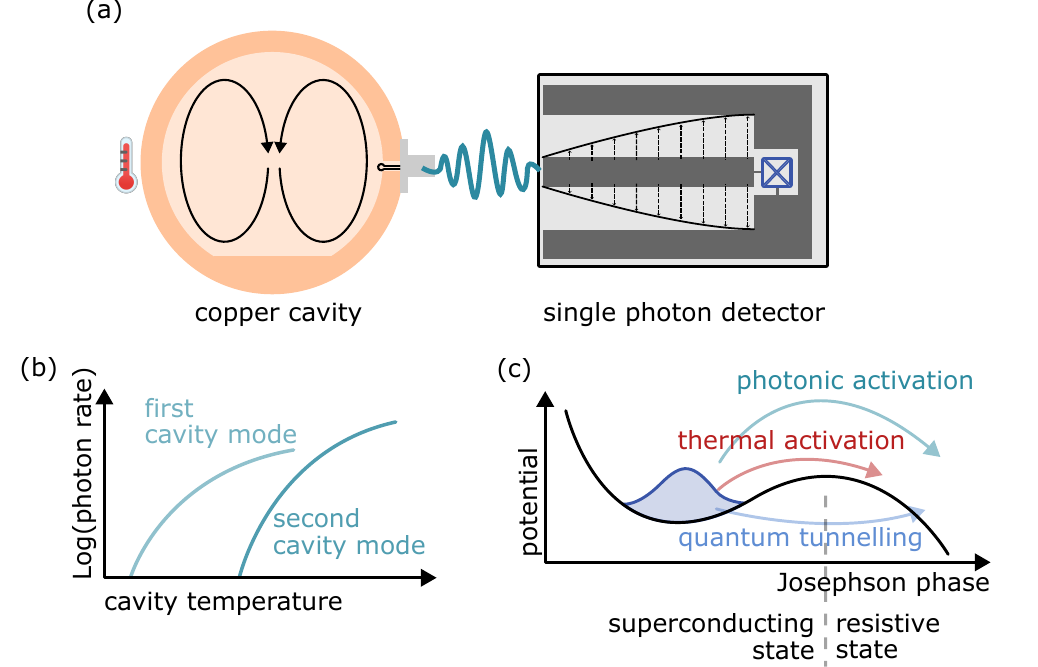}
    \caption{\textbf{Description of the experimental scheme.} a) A microwave cavity is connected to the photon detector, where the former acts as a source of thermal photons and the latter is formed by the Josephson junction, inserted into a coplanar waveguide adapter. b) Temperature dependence of the rate of thermal photons emitted by the cavity. With a corresponding temperature change of tens of millikelvin, the photon rate is varied by orders of magnitude. c) The tilted washboard potential describes the dynamics of a Josephson junction: the system is superconducting if the wavefunction is confined to a minimum, and resistive if it runs down the potential. The switching to the resistive state can be due to the arrival of a photon, or due to noise, appearing as thermal escape or quantum tunnelling processes, which therefore needs to be suppressed.}
    \label{fig:one}
\end{figure} 

The need of a microwave single photon detector (SPD) is witnessed by the field of Dark Matter particle searches, such as axions \cite{Sikivie2021Feb, Irastorza2018Sep}.
In particular, Dark Matter haloscopes \cite{PhysRevLett.124.101303, Kwon2021May, Adair2022Oct, Crescini2020May} rely on precision power measurements using low noise amplifiers, and the upgrade to photon detectors is utmost important \cite{Lamoreaux2013Aug, Barbieri2017Mar}, since it outperforms quantum limited amplifiers or quantum enhanced measurements \cite{Lamoreaux2013Aug, Backes2021Feb}. Especially at higher frequencies, i.\,e. above a few GHz, the standard quantum limit of linear amplification \cite{Lamoreaux2013Aug,Clerk2010Apr} drastically hinders the sensitivity of these measurements. 
The discovery of axions is a compelling argument, but the requirements needed for their detection are rather strict. The axionic signal is a stochastic emission of rare photons of frequency $f_\text{a}$, dependent on the unknown axion mass, in a narrow bandwith $10^{-6}f_\text{a}$ \cite{Sikivie2021Feb}. Therefore, a suitable photon detector should have a low dark count rate, wide spectral range, high efficiency, and be capable of continuous operation. Except for a few schemes circumventing some of these issues \cite{Albertinale2021,PhysRevApplied.21.014043}, these conditions are typically not fulfilled by existing microwave photon detectors, and only some recent experiments \cite{Dixit2021Apr,braggio2024quantumenhanced}, could tackle the issue.
The use of Josephson junctions as threshold detectors was proposed and analyzed in \cite{Kuzmin2018Jun}, but realised for detecting of high photon rates only \cite{Pankratov2022May}, while a detector of rare photons still poses an experimental challenge.

In this work, we demonstrate the detection of microwave single photons with high efficiency and sub-hertz dark count rate by using an underdamped Josephson junction (JJ). The photons are generated by the thermal emission of a bulk microwave resonator made of copper. The experimental scheme and principles of operation are shown in Fig.\,\ref{fig:one}. We engineer an experiment where the cavity design defines the frequency of the emitted photons, the rate of emission is adjusted from less than one photon per hour to thousands photons per second by varying the cavity temperature, and the photons are collected in the detector with an on-chip impedance matching line terminated by a JJ. 
When impinged by a photon, the current-biased junction switches to the resistive state making the detector click. 
The observed thermal photons depend exponentially on temperature and, when sourced from a single mode, follow a super-Poissonian statistics \cite{Fox2006,Kovalenko:23}, a unique signature of their nature. 

\section*{Results}
\begin{figure*}[!]
    \centering
    \includegraphics[width=\textwidth]{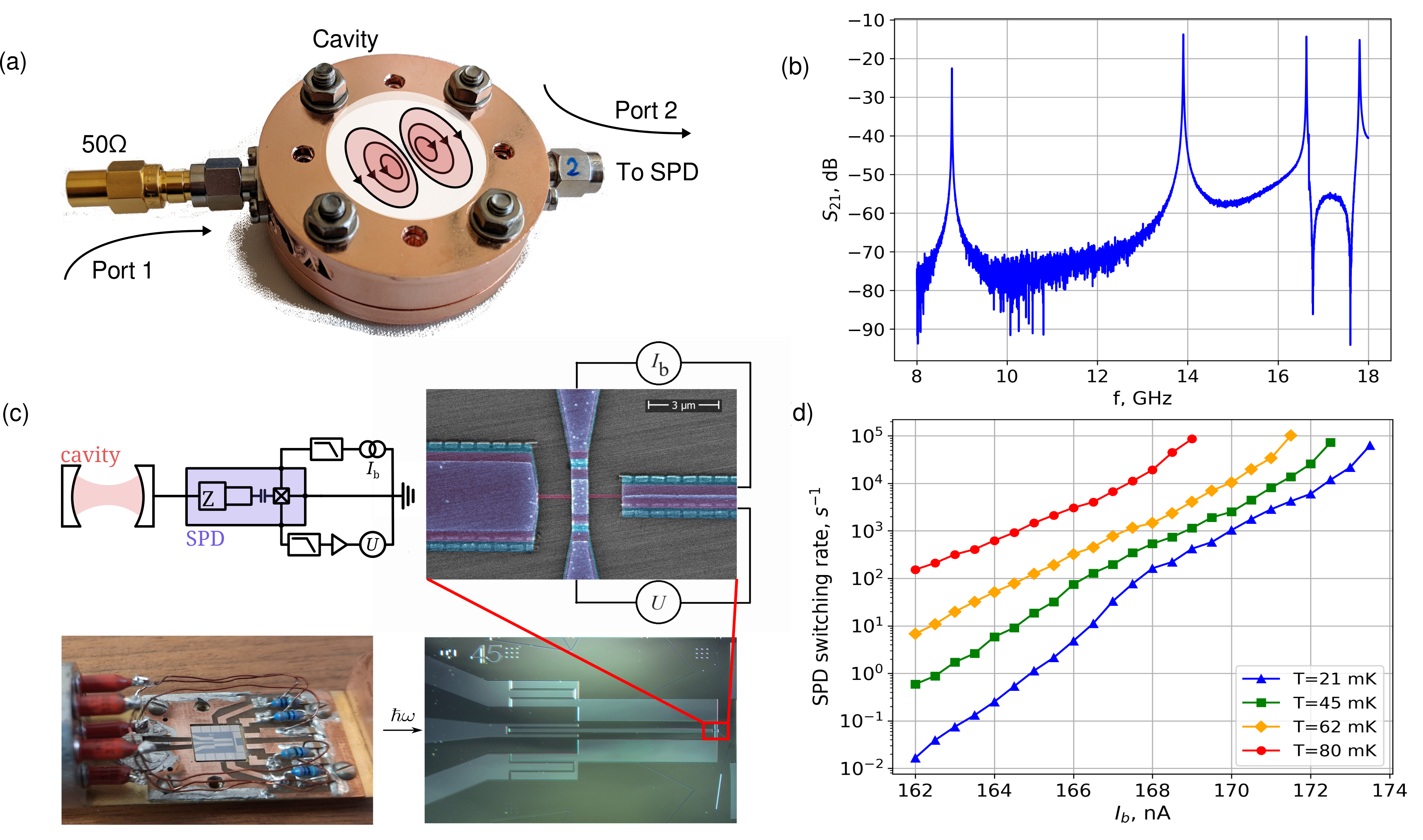}
    \caption{\textbf{Microwave cavity and photon detector used in this work.} (a) The cavity, critically coupled to the SPD through antenna port 2, while port 1 is very weakly coupled (used for characterisation purposes). At the centre of the cavity, a sketch of the 14\,GHz mode shape is shown. (b) The room temperature response of the cavity shows the resonant modes starting from the lower frequency mode at 8.81\,GHz. The comparison with resonances at 17 mK is shown in Methods. (c) A circuit diagram with optical images of the detector connected to the cavity antenna and mounted in the sample holder. SEM image of the Josephson junction with artificial colors: pink - bottom Al electrode, blue - top Al electrode, violet - overlapping areas. The feeding dc lines are also shown with bias current $I_b$ and measured voltage $U$. (d) The switching rates of the detector, measured at the base temperature 17\,mK, for various cavity temperatures from 21 to 80\,mK.}
    \label{fig:app}
\end{figure*}

\subsection*{Experiment concept}
The experiment is essentially composed of two parts: a photon source and a photon detector, shown in Fig.\,\ref{fig:app}.
The photon source of this experiment is a cylindrical cavity with the radius and length of about 3.0\,cm and 1.2\,cm, respectively, see Fig.\,\ref{fig:app}a. The body is oxygen-free high conductivity copper to ensure a proper thermalization and a quality factor of the order of $10^4$. The cavity is coupled to two SMA outputs (port 1, weakly coupled, and port 2, close to critically coupled) through two loop antennas. The measured $S_\text{21}$ transmission of the resonator is reported in Fig.\,\ref{fig:app}b, and shows the first four resonant modes.
The detector is an underdamped Al-AlO$_x$-Al Josephson junction directly connected to the cavity antenna and, upon the arrival of a photon, switches to the running state, producing a measurable voltage drop of hundreds of microvolts. The circuit diagram, the chip layout, the optical and scanning electron microscope (SEM) images of the coplanar line and Josephson junction are presented in Fig.\,\ref{fig:app}c, while the switching rate of the detector for several cavity temperatures is reported in Fig.\,\ref{fig:app}d. The on-chip coplanar line is designed as an impedance transformer from 50 to 200\,Ohm, but the matching efficiency is sufficiently high (i.\,e. above $60\%$) even for junctions with $R_N$ up to $2\,\mathrm{k}\Omega$ as shown by numerical simulations. The detector's bandwidth is approximately 3 GHz around 14 GHz, determined by the coplanar antenna. It also has a subband near 8.8 GHz, associated with the internal resonances of the Josephson junction sample.
The microwave cavity port 2 feeds the signal via a short coaxial cable, soldered to the PCB coplanar line, which is further bonded to the coplanar line of the chip. The T-filters for the bias and voltage lines are composed of feed-through capacitances (1.1 nF, red), on-plate resistors (500 Ohm, blue) and larger resistors outside the sample holder (10 kOhm, not shown). The output voltage is readout using room temperature AD745 amplifier and analog-to-digital converter. The SPD is located inside cryoperm and superconducting screens to minimise the magnetic noise and residual stray light, as shown in the Methods. 

The SPD is fabricated by shadow evaporation technique as Al-AlO$_x$-Al trilayer on a Si substrate, its area is $2.5\times 0.7\,\mu\text{m}^2$, the measured critical current at 17\,mK is $I_C=170\,\text{nA}$, and the normal state resistance is $R_N=1480\,\Omega$. The choice of the junction parameters is targeted to a reduced dark count rate and optimal quantum efficiency. The former requires a reduction of thermal activation and quantum tunneling events \cite{Kuzmin2018Jun}, while the latter is related to the junction critical current and damping. One can find a compromise between the efficiency and switching errors by utilization of the phase diffusion regime \cite{Pankratov2022May,Pankratov2024May}, see also Fig. 6 in Methods.
The sample presented hereafter has a critical current close to the optimal value for photon counting experiments \cite{Kuzmin2018Jun}, and was therefore chosen among others \cite{Pankratov2024May}, which were still capable of detecting single photons but with slightly worse performances. With the chosen parameters, the response time of the detector is of order inversed plasma frequency of the junction, so below 1 ns. However, the dead time of the detector, i.e., the time needed to restore the initial state after the switching, is primarily limited by the RC filters and readout electronics, and is $\sim$2–10 ms. 

An underdamped Josephson tunnel junction is well-known as a detector of a harmonic signal with a minimum detectable power down to femtowatts level, by measuring the photon-assisted tunneling (PAT) steps at the inverse branch of a current-voltage characteristic. In this work we use PAT steps to measure resonant curves of the cavity at low temperatures, as detailed in the Methods, to then demonstrate that the Josephson junction also works as a microwave single photon detector when its bias current $I$ is slightly lower than the critical current $I_C$. As outlined in Fig.\,\ref{fig:one}, the dynamics of a Josephson junction can be treated as evolution of a phase particle in a tilted washboard potential. An absorbed photon causes a current pulse $\Delta I$ through the junction, overcoming the threshold, and thus leading to the appearance of measurable resistive state with a finite voltage of about 0.4\,mV.
In the Josephson circuit, the energy $h f$ of the incoming photon is split between the energy of the supercurrent $E_s$ stored in the tank circuit and the energy $E_d$ dissipated in the subgap resistor $R_{qp}$, approaching $R_N$ close to zero voltage, according to quasiparticle IV curve of a tunnel Josephson junction \cite{Barone1982}. Assuming the simple relation between $E_s$ and the pulse amplitude $\Delta I$ \cite{Kuzmin2018Jun}, $E_s = L_\mathrm{JJ} {\Delta I}^2 / 2$, and taking into account that in parallel RLC-resonant circuit
    $Q = R_N \left( C / L_\mathrm{JJ} \right)^{1/2}$, $E_s / E_d = Q / 2 \pi,$
the current increase due to the photon can be estimated as $\Delta I = \sqrt{2 h f L^{-1}_\mathrm{JJ}(1 + 2 \pi / Q)^{-1}} $, 
where the junction inductance $L_\mathrm{JJ}$ depends on the bias current $I$ as $L_{JJ} = \hbar/(2e)(I^2_C - I^2)^{-1/2}$. Using the above parameters, the single photon current pulse $\Delta I$ is estimated about 60\,nA.

The average number of thermal photons in a cavity mode follows the Planck distribution and depends on the modes' resonance frequency, loaded quality factor, and temperature. Dissipation regulates the photon lifetime, giving rise to an emission rate that can be controlled by coupling an antenna to the mode \cite{Pozar2011}.
On first approximation, the cavity photon rate $r_\text{c}(T)$ measured with an SPD can be calculated considering $n$ modes as
\begin{equation}
    r_\text{c}(T) = \sum\limits_{i=1}^n \frac{\eta_i}{\tau_i}\frac{1}{e^{h f_i/k_\mathrm{B}T} - 1}
\label{photon_rate}
\end{equation}
where $f_i$, $\tau_{i}$ are the frequencies and the lifetimes of the cavity modes' photons, respectively; $h$ is the Planck constant, $k_\mathrm{B}$ is the Boltzmann constant and $T$ is the temperature. The photon lifetime $\tau_{i}=Q_i/(2\pi f_i)$ is extracted from the quality factor $Q_i$ of the mode and its central frequency $f_i$. The parameter $\eta_{i}$ is the ratio between the thermally available photons of the $i$-th cavity mode and the detected ones. It depends on the cavity antenna coupling and on the detector quantum efficiency. The fact that the antenna is critically coupled to the cavity means that only a half of available photons exit the cavity and go to the detector.
Thanks to the exponential dependence of the rate on the $h f_i/k_\mathrm{B}T$ ratio, at low enough temperatures only a limited number of modes significantly contribute to the rate, allowing us to neglect the effect of higher order resonances.

\subsection*{Single photon detector design and properties}
\begin{figure*}
    \centering
    \includegraphics[width=1.\textwidth]{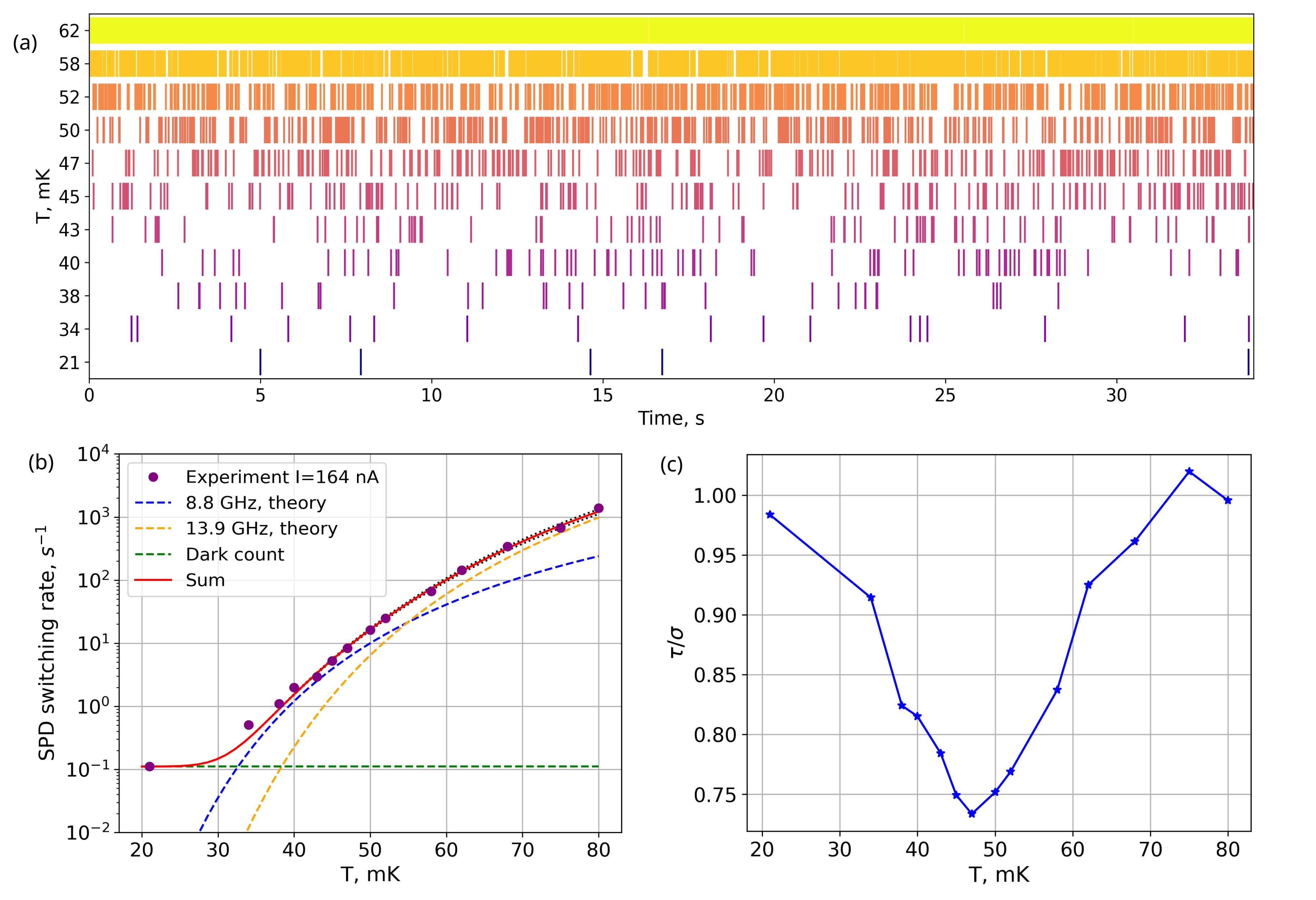}        
    \caption{\textbf{Single microwave photon detection.} a) An event plot of the recorded data is shown at various temperatures. b) The SPD switching rate, fitted with the theoretical photon rate, using formula \eqref{photon_rate}. The error bars do not exceed the dot sizes due to 5000 averages. The data are fitted with the expected cavity photon rate $r_\text{c}$ \eqref{photon_rate} (red solid curve) plus a dark count rate $r_\text{DC}$ (green dashed line) to extract the efficiency and noise floor of the detector. The black dotted curves show $\pm0.05$ $\eta_2$ spread of the fitting. The values of quantum efficiency $\eta$ reach $\sim 0.0125$ and $0.45\pm0.05$ for 8.81 GHz and 13.9 GHz modes, respectively.  c) The $\tau$ to $\sigma$ ratio vs cavity temperature, showing deviation from Poissonian statistics.}
    \label{fig:phrate}
\end{figure*}

After a preliminary room-temperature spectroscopic characterization, the cavity was mounted on the mixing chamber plate of a dilution refrigerator, and connected to the SPD.
The cavity spectroscopy was repeated at 17\,mK by sending a microwave tone to the cavity via port 1 and observing the first three cavity resonances as PAT steps in the current-voltage characteristic of the junction \cite{Pankratov2022May,stanisavljeviс2023nearideal}.
The results of these measurements are shown in Methods.

In Fig.\,\ref{fig:app}d, the SPD is characterised by measuring its mean switching rate versus a bias current for various cavity temperatures. To this end, the temperature of the cold plate of a dilution fridge was fixed at about 17\,mK, while the cavity, only weakly thermally coupled to the plate, was heated by a resistor to a temperature of 21 to 80\,mK, precisely controlled using a SQUID noise thermometer. The mean switching rate coincides with the SPD dark count rate at the lowest cavity temperature of 21 mK in absence of thermal photons, while it is defined by the photon rate at a large cavity temperature. One can see that the switching rate, starting from 0.01\,Hz at the smallest bias current, increases by four orders of magnitude when heating cavity up to 80\,mK, demonstrating the efficient response to incoming photons. It was checked that the dark count rate nearly does not change at the same cavity temperature variation if the sample is disconnected from the cavity, and also that the response is heavily suppressed if the cavity is connected to the SPD by the port 1.

\begin{figure*}
    \centering
    \includegraphics[width=0.8\textwidth]{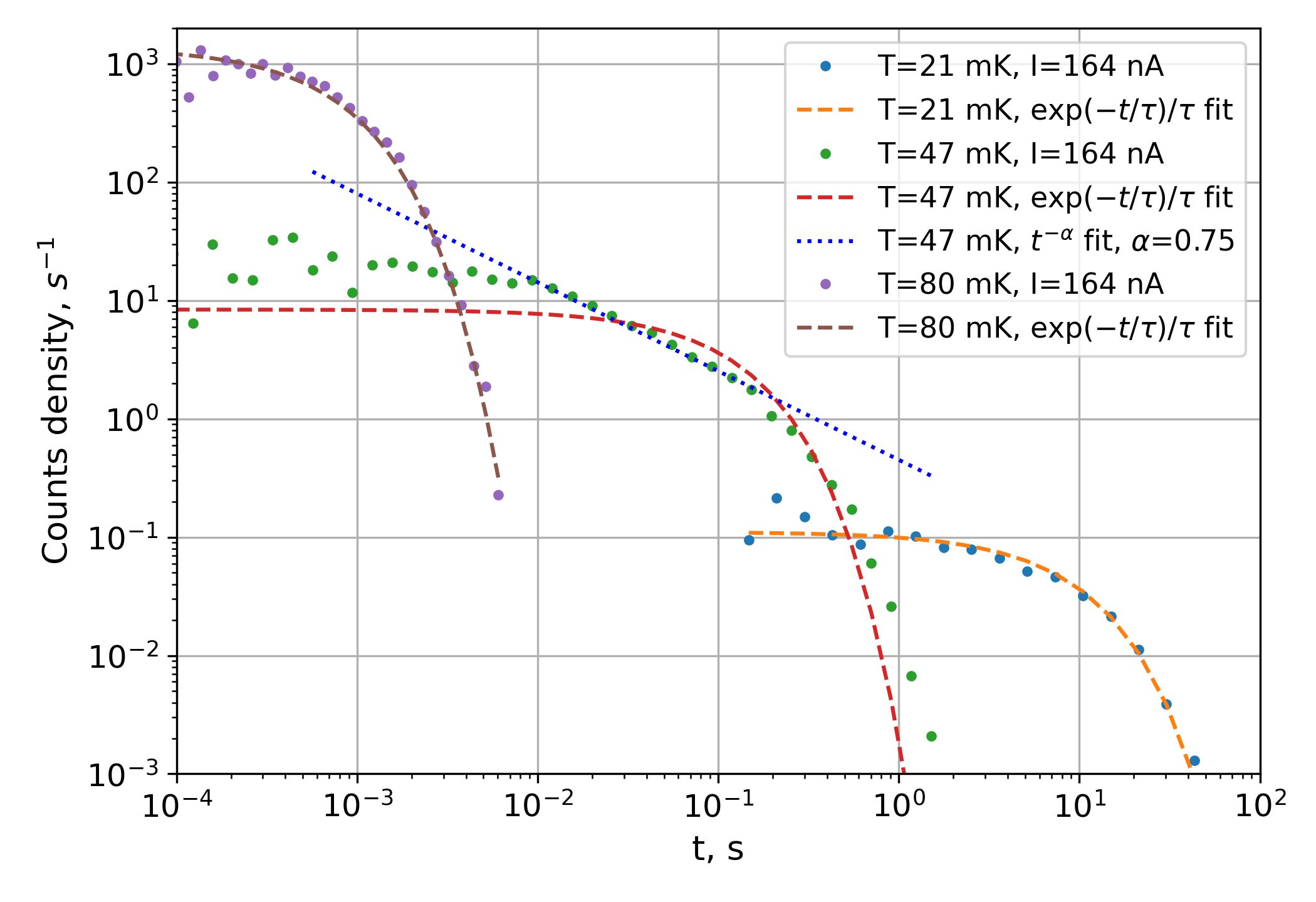}
    \caption{\textbf{Distribution of times between sequential switchings of the SPD at three various temperatures.} The experimental data (dots) are fitted by $\exp(-t/\tau)/\tau$ (dashed curves) without any fitting parameters with the mean switching time $\tau$ directly taken from experimental data. Good agreement between fitting and experiment is observed at 21 mK and 80 mK cavity temperatures. One can note super-Poissonian distribution at intermediate cavity temperature 47\,mK. }
    \label{fig:lt_dist}
\end{figure*}

To observe the variation of the photon rate and characterise the quantum efficiency and dark count rate of the detector, we measured the temperature dependence of the switching events. 
In Fig.\,\ref{fig:phrate}a  we present an event plot of the recorded data at various temperatures, from 21 to 62\,mK, showing an exponential increase of the switching rate. Fig.\,\ref{fig:phrate}b displays the photon rate vs temperature and its fit. We fit the data with the expected cavity photon rate $r_\text{c}(T)$ in Eq.\,\eqref{photon_rate} plus a dark count rate $r_\text{DC}=1/\tau_0$, which is the inverse of the mean switching time $\tau_0$ taken from the lowest temperature data, to extract the efficiency and noise floor of the detector. For this analysis, the quality factors and resonance frequencies are fixed to the values measured using PAT steps, see Methods. In particular, at $f_1 = 8.81\,\mathrm{GHz}$, $Q_1\simeq 7340$, and at $f_2 = 13.95\,\mathrm{GHz}$, $Q_2 \simeq 4650$. Fig.\,\ref{fig:phrate}b shows a good agreement between the expected cavity photon rate and the experiment.
From this data, we verify that only the first two modes contribute to the detector response at these temperatures, while the higher frequency modes are irrelevant. The detection efficiency $\eta_2$ is as high as $45\pm5$\% for the mode $f_2$, with a dark count rate of 0.1\,Hz. 
In this experiment, the estimated efficiency is a convolution between the antenna coupling and the detector efficiency, and is therefore a lower limit.
For instance, $\eta_1\simeq1\%$ is mainly due to the weak coupling of the antenna to the mode $f_1$, as can be deduced from Fig.\,\ref{fig:app}b, as well as SPD coplanar antenna selectivity, aimed to efficiently receive 14\,GHz photons.
The choice of the antennas' couplings is tailored to swap the main contribution to $r_c(T)$ from $f_1$ to $f_2$ at a temperature of about $50\,\mathrm{mK}$, therefore forming a peculiar cavity response within the dynamics of the detector, as shown in Fig.\,\ref{fig:phrate}b.

Since a signature of single mode thermal photons is their super-Poissonian statistics \cite{Fox2006}, let us study the ratio of the SPD mean switching time $\tau$ to its standard deviation $\sigma$. It is known that in the case of Poissonian statistics, which is a natural statistics of thermal or quantum dark counts of our detector, this ratio should be equal to unity \cite{Malakhov2002Jan}. In Fig.\,\ref{fig:phrate}c the $\tau / \sigma$ ratio vs temperature is shown. One can see that at low and high temperatures $\tau / \sigma$ is close to unity, proving the Poissonian statistics. At intermediate temperatures, $\tau / \sigma$ is significantly lower than unity, reaching the minimum value of 0.73 at 47\,mK. This is the evidence of the super-Poissonian distribution of photons, also referred to as photon bunching. Fig.\,\ref{fig:lt_dist} illustrates how it is reflected in the probability density distributions.

The time intervals between the events are presented as  switching time distributions in Fig.\,\ref{fig:lt_dist} and fitted with an exponential distribution (dashed curves). This probability density for noise-induced escapes (or tunneling) across (through) the barrier should represent Poissonian distribution in the form $w(t)=\exp(-t/\tau)/\tau$ \cite{Malakhov2002Jan}. One can see that the curve for 21\,mK is fitted by the above exponential dependence without any fitting parameter for $\tau=9.026\,\text{s}$. The curve for 80\,mK is also well-fitted by exponential dependence with $\tau=7.214\,\text{ms}$. In all cases, $\tau$ is taken directly from the experimental data as the mean time interval between the events. At the same time, at 47\,mK, the switching time distribution is better fitted by the power dependence $t^{-\alpha}$ with $\alpha=0.75$ (dotted curve) than by the exponential dependence with $\tau=0.119\,\text{s}$, at least for intermediate switching times. This is the evidence of quantum chaos \cite{Yusipov2020Feb} as natural statistics of thermal photons \cite{Fox2006}, and is further discussed in the Methods. The change from Poissonian distribution at low temperatures to super-Poissonian at intermediate and back to Poissonian at high temperatures can be explained in the following way. At low temperatures, when the thermal photons in the cavity are very rare, the Poissonian statistics is the internal statistics of the detector as, e.g., for noise-induced escapes across a potential barrier \cite{Malakhov2002Jan}. With temperature increase, the contribution of thermal photons from a single 8.81 GHz mode starts to dominate, that, as argued in \cite{Fox2006}, should demonstrate the super-Poissonian statistics (see the mode contribution vs temperature in Fig.\,\ref{fig:phrate}b). With further increase of the temperature, the main 14\,GHz mode starts to compete with the first one (their contributions become equal at 54 mK), that leads to Poissonization of the thermal photon statistics, according to \cite{Fox2006}. To our knowledge, this is the first direct observation of the super-Poissonian statistics of microwave thermal photons.

\section*{Discussion}
In this work we describe and operate a microwave single photon detector used to observe the clicking and the statistics of thermal photons emitted by a resonator at ultracryogenic temperatures.
The Josephson junction-based sensor shows sub-Hz dark count rate and high efficiency, combined with a reduced operation complexity compared to qubit-based designs \cite{Dixit2021Apr,Albertinale2021}. 
The SPD was used to observe single microwave photons emitted by a single-mode and multi-mode thermal source, i.\,e. a copper cavity. Although widely accepted \cite{Fox2006,PhysRevLett.16.1012,Boitier2009}, an experimental demonstration of the super-Poissonian statistics of thermal microwave photons was, to our knowledge, missing. 
The measured photon rate is consistent with thermal emission from the microwave cavity modes, and closely follows the expected temperature dependence.
The photon distribution displays a clear reduction of the mean to standard deviation ratio when the light is emitted mainly by one mode, which instead remains unitary for dark count and multi-mode emission.
The dynamical range of the detector extends from its dark count rate to the kHz range and is currently limited by the acquisition electronics' speed. The detector dynamics was not further investigated, and will be the subject of future works.
The SPD efficiency can be better characterised with an on-demand microwave photon source \cite{Bozyigit2011,Lang2011Jul,Peng2016Aug,Zhou2020Mar}, to disentangle it from other systematic uncertainties. Future efforts to improve the SPD aim mainly at improving its dark count rate. For instance, a junction with lower $I_C$ could enter the phase diffusion regime, which is expected to increase its lifetime, and thereby reduce the dark count rate \cite{Pankratov2022May}. Advancing the screening and filtering of the setup is also foreseen to improve the detector performances.
In perspective, this work paves the way to a deeper examination of thermal light's quantum properties, and to their use to study quantum chaos \cite{Yusipov2020Feb}, ghost imaging \cite{PhysRevLett.93.093602,Ou_2007,PhysRevA.79.053844}, and more \cite{Fox2006}.
The device itself finds numerous applications in the field of precision physics beyond the Standard Model, as for instance Axion searches \cite{Lamoreaux2013Aug,Barbieri2017Mar}. On the one hand, the performance requirements needed to drastically improve current Axion experiments are already met by the current device, on the other hand, these apparata typically include strong magnetic field and long operation times, posing a challenge to the screening and long-term-stability of the detector. Nevertheless, the implementation of single microwave photon detectors such as the one presented here fosters the emergence of next-generation beyond the Standard Model experiments.

\section*{Acknowledgments}
The work is supported by the Russian Science Foundation (Project No.~\mbox{19-79-10170}). This project is also supported by the Italian Institute of Nuclear Physics (INFN) within the QUAX experiment. The authors wish to thank G. Carugno for triggering the interest on photon detectors and for the multiple advice in the course of this work, A.A. Yablokov for writing an automated Python-based software, and I.V. Rakut' for sample holder fabrication.

\section*{Author contributions}
The general idea of the work was suggested by L.S.K., N.C., A.V.G. and A.L.P. The sample fabrication was performed by A.V.G and A.V.B. A.L.P. ran the experiment with support of A.V.G., A.V.C. and L.S.R. Data were analyzed by N.C., A.L.P., A.V.G. and A.V.C. A.L.P., A.V.G., L.S.R. and A.V.B. designed and assembled the experiment. The cavity was designed by N.C., and tested by N.C., A.L.P., A.V.G. and A.V.C. All authors, led by N.C. and A.L.P., contributed to manuscript with figures created by N.C., A.V.G. and A.V.C.

\backmatter


\section*{Methods}

\subsection*{Sample fabrication}
The fabrication of a Josephson junction based single photon detector requires a two-layer process. In the first  Ti/Au/Pd multilayer, the contact pads, the antenna, and dc wires were fabricated at the nanofabrication facility of Chalmers university using laser-writer and lift-off technology. In the second layer, we patterned the superconductor-insulator-superconductor (SIS) tunnel junctions with the electron beam lithography. Thereafter, the SIS junction structures made of aluminum were deposited at NNSTU using a well-established shadow evaporation technique \cite{Gordeeva2020Dec,Moskalev2023Mar} without breaking the vacuum. This technology is typically used for fabrication of superconducting qubits that ensures long time operation of the samples.

\subsection*{Heating the resonator}
\begin{figure}[!]
    \centering
    \includegraphics[width=0.5\textwidth]{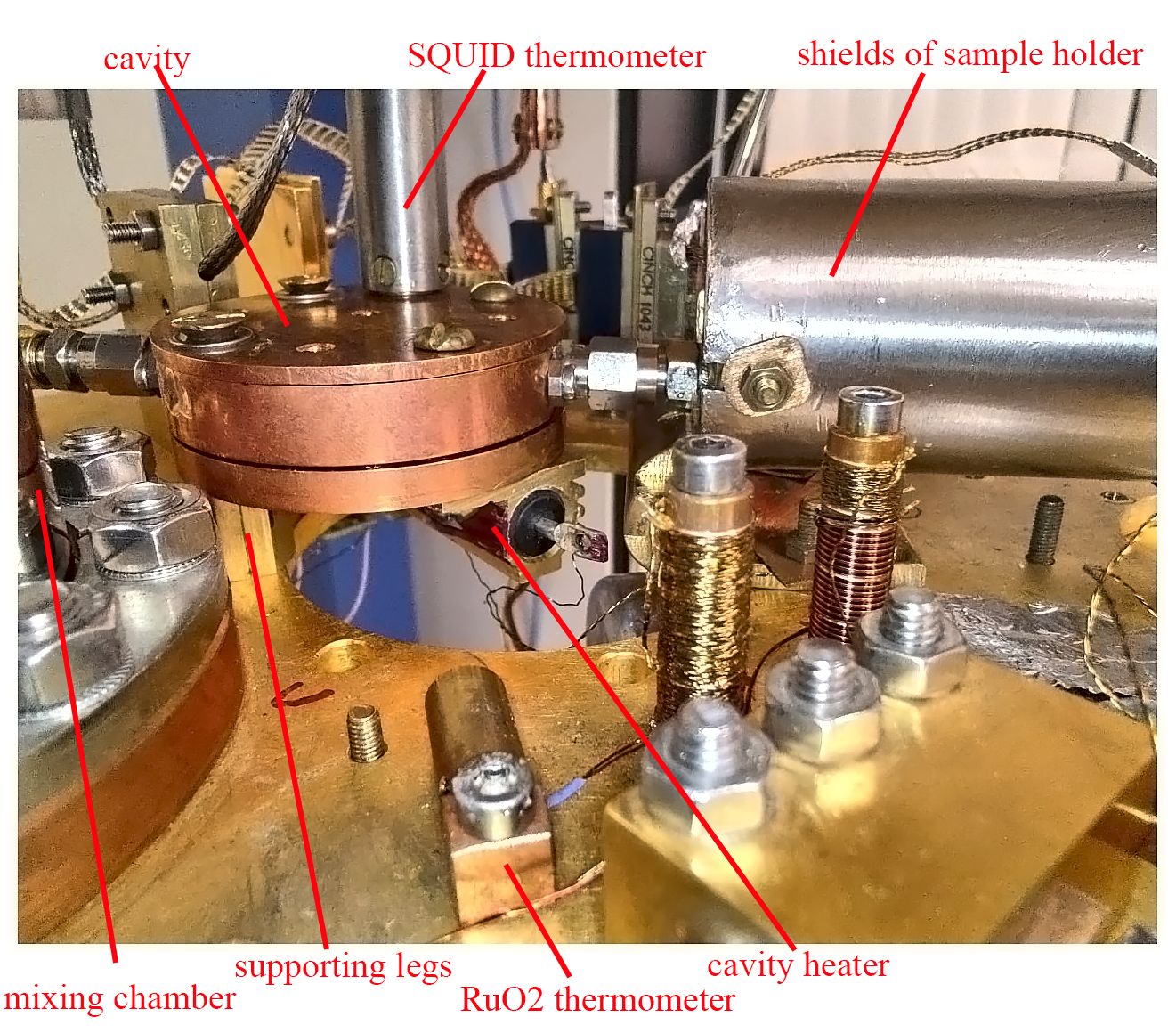}     
    \caption{\textbf{Picture of the 10\,mK plate of the dilution fridge.} One can see the sample holder in the cryoperm screen and the copper cavity with the SQUID thermometer on top and the heating resistor under the bottom.}
    \label{fig:methods_cav}
\end{figure}

Fig.\,\ref{fig:methods_cav} shows the setup for mounting the cavity on the 10\,mK plate of a dilution refrigerator with a weak thermal coupling. A SQUID thermometer is installed on top of the resonator for precise temperature control. A 5$\,\Omega$ heating resistor is installed at the cavity bottom. Measurements of the response of the SPD to the signal from the resonator are performed for heating in the temperature range from 21\,mK to 80\,mK, which corresponds to the voltage range of the heating resistor from 0\,mV to 28\,mV. The copper basement for the SPD is made of a single piece of copper directly anchored to the mixing chamber plate and the heat exchange between the SPD and the cavity goes mostly through the SMA connector. The measurements of the SPD switching time were carried out with Python-based software package, allowing precise feedback of the cavity and cryostat plate temperatures. 
\begin{figure*}[!]
    \centering
    \includegraphics[width=1.\textwidth]{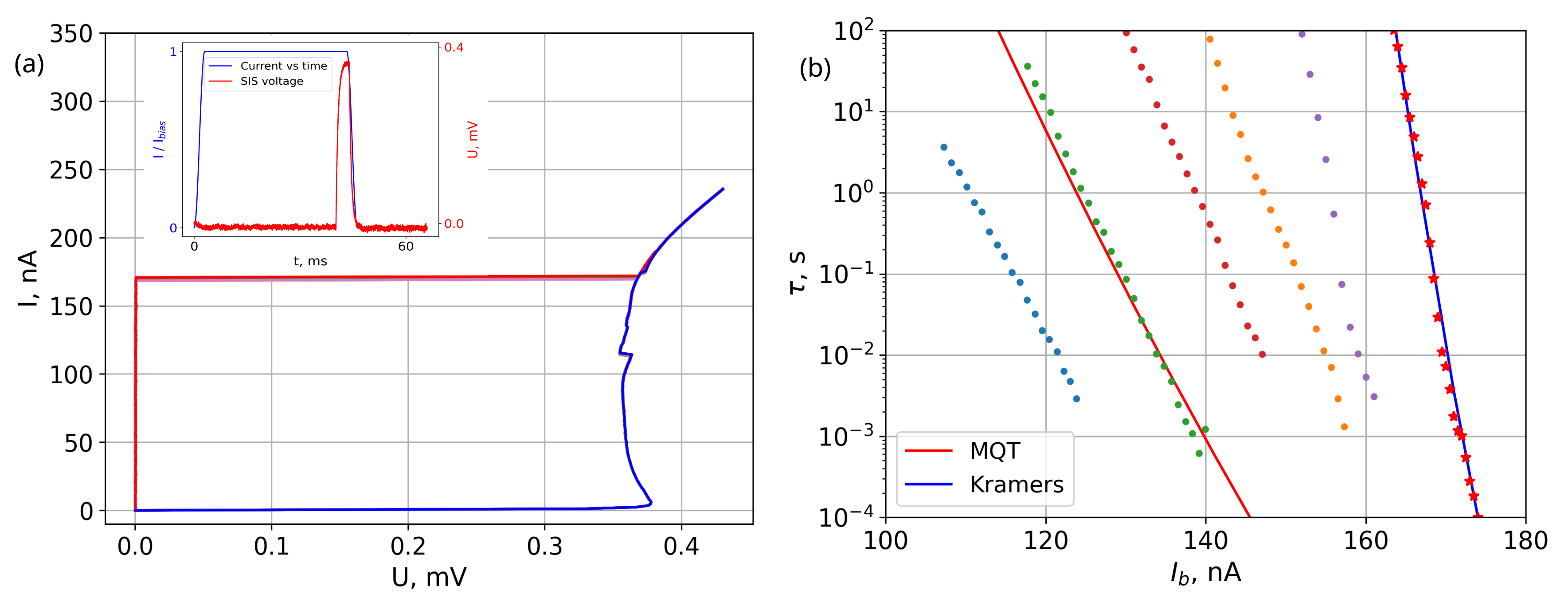}
    \caption{\textbf{Preliminary characterisation of the detector.} a) The current-voltage characteristics of the sample at 17\,mK, sequentially measured 10 times to show the stability of measurement setup. Inset: example of a single shot measurement with bias current rise and fall times of 3 ms each. b) The mean switching times of the SPD, indicating drastic decrease of dark counts during several months progress of filtering improvement (dots - experiment, solid curves - theory). The error bars do not exceed the dot sizes.}
    \label{fig:methods_iv_tau}
\end{figure*}

\subsection*{Shielding of the detector}
The shielding of the setup from the environment is straightforward. The detector is readout with dc lines, filtered by the T-filters already described in the main text. Photons are emitted and measured inside the compact volume of the cavity and the detector holder without the need of any microwave cable, thereby ensuring a strong decoupling from the environment. External magnetic fields around the detector are screened by two-layer cryoperm and superconducting screen. Our experimental setup shows no sensitivity to either the Earth's magnetic field or nearby magnets. 

A typical current-voltage characteristic of the sample SPC-73 sequentially measured 10 times is shown in Fig. \,\ref{fig:methods_iv_tau}a to visualize the stability of the measurement setup. The measured critical current $I_C=\mathrm{170\,nA}$ at 17\,mK is very close to the theoretical limit of 197\,nA according to the Bardeen–Cooper–Schrieffer (BCS) theory for the idealized case of the absence of both thermal and quantum noise. An example of a single shot measurement with bias current rise and fall times of 3 ms each and the switching time of 42 ms is shown in the inset of Fig. \,\ref{fig:methods_iv_tau}a.

The mean switching time of the sample SPC-73 is shown in Fig.\,\ref{fig:methods_iv_tau}b, and indicates the drastic decrease of dark counts after many iterations of filtering and screening of the setup (from left to right). Starting from first experiments, our goal was to improve filtering to reach the theoretical limit, given by the tunneling time formula \cite{Martinis1988Aug,Oelsner2013Sep,Golubev2021Jul}. In Fig.\,\ref{fig:methods_iv_tau}b the tunneling time from the potential well minimum \cite{Golubev2021Jul} is shown by the red solid curve. After approaching this theoretical limit, additional efforts were given to cancel the residual stray light \cite{Barends2011Sep}. Step-by-step, we have reached the rightmost curve (red dots), thus significantly exceeding the tunneling time theoretical limit \cite{Martinis1988Aug,Oelsner2013Sep,Golubev2021Jul}. This result means that we have, presumably, entered into the quantum phase diffusion regime, which is not described by the standard formulas, but using which the dark count time can be increased by several orders of magnitude \cite{Pankratov2022May}. Interestingly, the rightmost curve, shown by red dots, can be well fitted by the classical Kramers' time \cite{Malakhov2002Jan} (blue curve) for realistic temperature 18\,mK and fitting critical current value 191\,nA, which is higher than 170\,nA, measured in experiment, but roughly corresponds to BCS theoretical limit for this sample.

\subsection*{Fitting the cavity emission rates and measuring its resonances by photon-assisted tunneling steps}
\begin{figure*}[!]
    \centering
    \includegraphics[width=1.\textwidth]{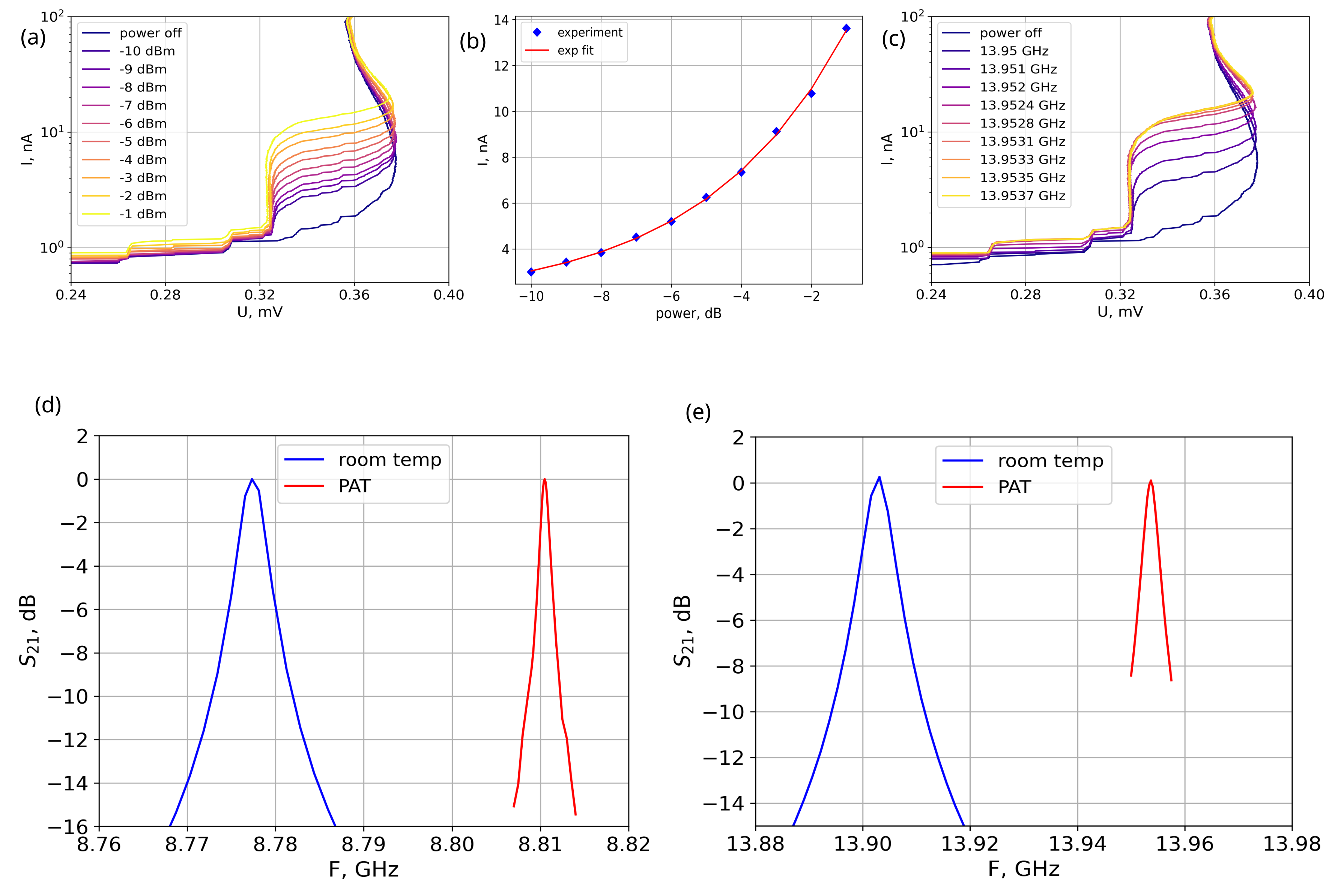}
    \caption{\textbf{Measurements of cavity resonances by photon-assisted tunneling steps of the SPD.} a) IV curves at various incident radiation powers and constant frequency of 13.9537 GHz; b) Calibration curve to translate the position of current step to dB; c) IV curves at various frequencies of incoming radiation and constant power; d,e) Comparison of frequency response of the cavity measured at room temperature by VNA (blue) and at 17\,mK measured by photon-assisted tunneling steps in SPD (red). d) Resonance at 8.8\,GHz, e) resonance at 13.9\,GHz. Both dependences are normalized to the same level so that their maxima are at 0\,dB. Blue curves are actually the enlarged curves from Fig. \ref{fig:app} (b).}
    \label{fig:methods_PAT}
\end{figure*}

In the case where the SPD is connected to a resonator, we obtain a significant increase in the switching rate if the resonator is heated, see Fig. \ref{fig:app}d. When the detector is disconnected from the resonator, the switching rate only slightly increases due to the small influence of the resonator heating on the sample holder at the maximal heating voltage. 

In order to correctly fit the measured photon rate with the theory, see Fig.\,\ref{fig:phrate}b, we must properly characterise the cavity quality factors of the first two modes as well as their central frequencies and substitute these values into Eq.\,\eqref{photon_rate}.
The resonator has four resonant frequencies: 8.81\,GHz; 13.95\,GHz; 16.67\,GHz and 17.85\,GHz, but the rate of photon emission at frequencies 16.67\,GHz and 17.85\,GHz is more than one order of magnitude lower than at the other two frequencies. Therefore, when determining the efficiency of the counter, only modes at 8.81\,GHz and 13.95\,GHz were taken into account. The switching efficiency of the detector consists in three components: the emission efficiency of the antenna, the matching of the resonator with the detector, and the switching efficiency of the detector due to photons. Moving away from the optimal bias current point, the switching efficiency at 8.81\,GHz drops off quickly, while the efficiency of 13.95\,GHz photon detection, for which the detector was elaborated, drops more slowly and still remains reliable. 
\begin{figure*}[!]
    \centering
    \includegraphics[width=1.\textwidth]{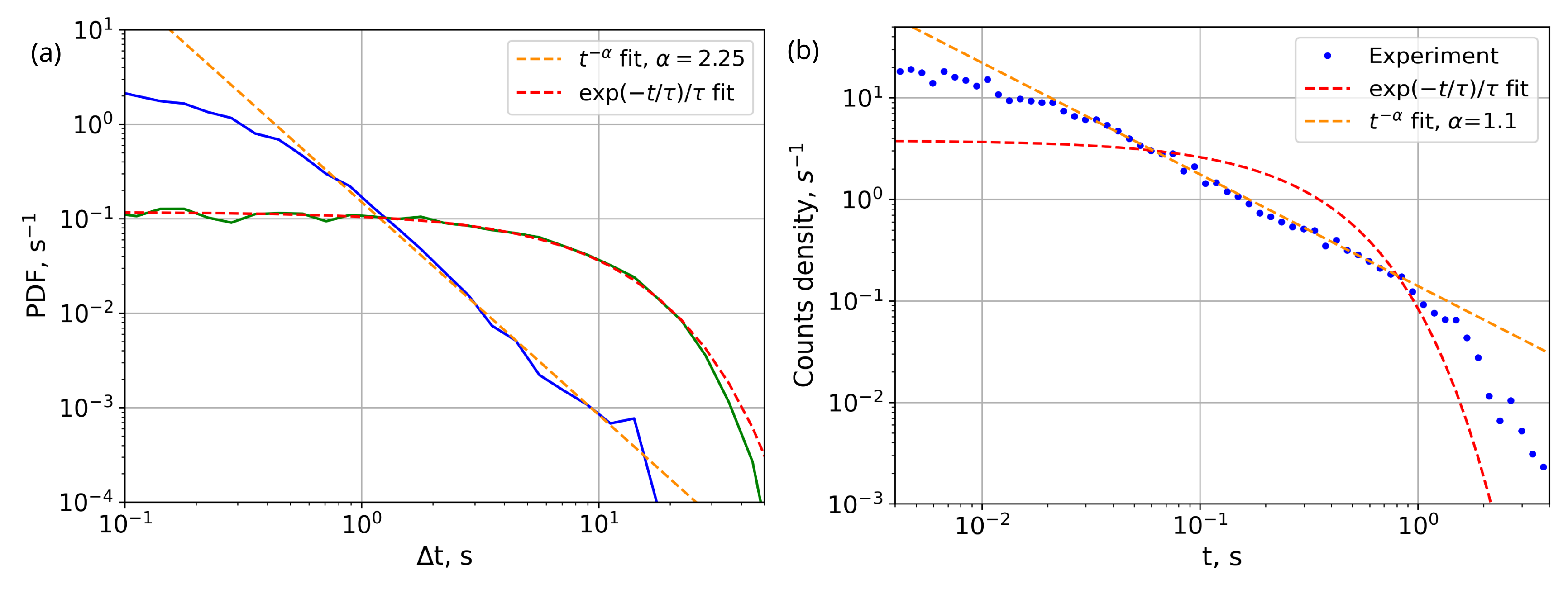}
        \caption{\textbf{Comparison of Poissonian and super-Poissonian distributions.} a) Numerically computed distributions according to \cite{Yusipov2020Feb}. b) Measured switching time distribution (blue dots) of thermal photons from a cavity representing super-Poissonian distribution, fitted with exponent (red dashed curve) and $t^{-\alpha}$ dependence.}
    \label{fig:methods_lt_hist}
\end{figure*}

Since we did not have an opportunity to directly measure frequency response of the cavity at low temperatures using a vector network analyzer (VNA), we have performed a similar measurements by using the photon-assisted tunneling steps of the SPD's IV curves \cite{Tien1963Jan}. When the voltage at the junction is slightly lower than the doubled gap voltage, a single electron (quasiparticle) is able to tunnel from one side of the junction to another one. It becomes possible when $eV + h f > 2\Delta$ with $V, f, \Delta$ being voltage, emission frequency and superconducting energy gap, respectively. This effect appears in the IV curve of the junction as PAT step. The position of this step depends on the power of the incident signal while the voltage depends on its frequency.

The method for measuring the frequency response of the cavity with the use of PAT steps is as follows. First of all we measure a series of IV curves at a frequency close to a resonance with various microwave power values (Fig.~\ref{fig:methods_PAT}a). Then we plot the calibration curve which shows the dependence of the current step position on supplied microwave power in dB at a certain voltage on the SPD (Fig.~\ref{fig:methods_PAT}b, blue diamonds). We make an exponential fit of the calibration curve and thus have the dependence of the incident power on the current step position at any value between the measured points (Fig.~\ref{fig:methods_PAT}b, red curve). Then, finally, we measure IV curves at a constant supplied microwave power with the frequencies varying in the vicinity of the cavity resonance (Fig.~\ref{fig:methods_PAT}c). Using the calibration curve obtained at the previous step we transform the positions of the current step into the value of the incident power in dB. Since we make such measurements near the resonant frequency, the incident power strongly depends on the frequency of microwave signal. The frequency response of the cavity near 8.8\,GHz (d) and 13.9\,GHz (e) measured by PAT steps (red curves) are presented in Figs.\,\ref{fig:methods_PAT}d,e. For comparison, we plot room temperature frequency response of the cavity, measured with a VNA. We see that the quality factor at low temperature 17\,mK increases, as expected, with simultaneous shift of resonant frequencies to slightly higher values. Thus, at low temperatures the quality factor of the 8.81\,GHz mode reaches $Q_1=7340$, and the one of the 13.9\,GHz mode is $Q_2=4650$. The measured values of quality factors and central frequencies are used for fitting of the measured photon rate with the theory in Eq.\,\eqref{photon_rate}, shown in Fig.\,\ref{fig:phrate}b.

\subsection*{Analysis of the switching statistics}
The observation of super-Poissonian statistics of photons was proposed to be used as a signature of quantum chaos \cite{Yusipov2020Feb}. Figure\,\ref{fig:methods_lt_hist}a shows the results of quantum chaos modeling \cite{Yusipov2020Feb}, where the authors propose to use photon arrival histograms as a tool for observing quantum chaos. The authors compare the results of numerical simulations, and show that the statistics change from Poissonian for regular dynamics to super-Poissonian in the case of the quantum chaos appearance. This is exactly the distribution observed in our experiment, shown in Figure\,\ref{fig:methods_lt_hist}b, supporting the slogan: quantum chaos -- nonlinear linearity. Here, we should add, that while the cavity represents completely linear resonator, the total system is not linear, since the resonator is weakly coupled to the Josephson junction detector, which is a strongly nonlinear element.

\section*{Data Availability}
The data that support the findings of this study are available from A.L. Pankratov upon reasonable request.

\section*{Competing Interests}
The authors declare no competing interests.

\end{document}